\documentclass[aps,prl,preprint,groupedaddress]{revtex4}

\usepackage{graphicx}
\usepackage{docs}
\usepackage{bm}

\begin{document}


\preprint{Preprint UCLA-PBPL0210 Submitted to Physical Review Letters}
\title{Teravolt-per-meter plasma wakefields from low-charge, femtosecond electron beams}


\author{J.~B.~Rosenzweig$^*$ }
\author{G.~Andonian$^*$ }
\author{P.~Bucksbaum$^\dagger$ }
\author{M.~Ferrario$^1$ }
\author{S.~Full$^*$ }
\author{A.~Fukusawa$^*$ }
\author{E.~Hemsing$^*$ }
\author{M.~Hogan$^\dagger$ }
\author{P.~Krejcik$^\dagger$ }
\author{P.~Muggli$^2$ }
\author{G.~Marcus$^*$ }
\author{A.~Marinelli$^*$ }
\author{P.~Musumeci$^*$ }
\author{B.~O'Shea$^*$ }
\author{C.~Pellegrini$^*$ }
\author{D.~Schiller$^*$ }
\author{G.~Travish$^*$ }
\affiliation{$^*$Department of Physics and Astronomy, University of 
California Los Angeles, Los Angeles, California 90095, USA}
\affiliation{$^\dagger$Stanford Linear Accelerator Center, Menlo Park, CA}
\affiliation{$^1$ Istituto Nazionale di Fisica Nucleare Ð Laboratori Nazionali di Frascati, via Enrico Fermi 40, Frascati (RM) Italy}
\affiliation{$^2$University of Southern California, Dept. of Engineering Physics, Los Angeles, CA}

\date{\today}

\begin{abstract}
Recent initiatives in ultra-short, GeV electron beam generation have focused on achieving sub-fs pulses for driving X-ray free-electron lasers (FELs) in single-spike mode. This scheme employs very low charge beams, which may allow existing FEL injectors to produce few-100 as pulses, with high brightness. Towards this end, recent experiments at SLAC have produced $\sim$2~fs rms, low transverse emittance, 20~pC electron pulses. Here we examine use of such pulses to excite plasma wakefields exceeding 1~TV/m. We present a focusing scheme capable of producing $<$200~nm beam sizes, where the surface Coulomb fields are also $\sim$TV/m. These conditions access a new regime for high field atomic physics, allowing frontier experiments, including sub-fs plasma formation for wake excitation.
\end{abstract}


\maketitle

Use of low charge, in the pC range, has been recently proposed for enabling GeV-class beams to be compressed to the hundreds of attosecond level \cite{rosen-nima-08}. 
Further, these beams are predicted to have very low normalized transverse emittance $\epsilon_n$, and thus unprecedented brightness.  
This proposal was generated in the context of planning future X-ray self-amplified spontaneous emission free-electron lasers (SASE FELs \cite{bonifacio}). 
The scheme addresses two challenges in the X-ray SASE FEL: it breaches the fs frontier in X-ray pulse length, and it  allows single spike SASE FEL performance. 
Both properties permit exploitation of the revolutionary aspects of coherent X-ray FEL light, as one may resolve properties of atomic and molecular systems at the spatial and temporal scales relevant to electronic motion \cite{neutze}.

With the advantages of low charge operation well appreciated, initial experimental work has been undertaken at the Stanford Linear Coherent Light Source (LCLS) recently.  
In these initial tests, 20~pC beams were compressed to rms duration $\sigma_t \simeq 2$~fs, while achieving emittances  $\epsilon_{n,x(y)} = 0.14 (0.4)$~mm-mrad at an energy  $U_b=$~14~GeV. 
This beam, having very high current $I$ and low emittance $-$ and thus high brightness, $B=2I/\epsilon_{n,x}\epsilon_{n,y}$ $-$ is predicted to produce nm-wavelength FEL pulses in the single-spike regime \cite{ding}.  

Beams with such short duration can produce coherent electromagnetic excitations with frequency components up to  $\omega_{max} \simeq \sigma_t^{-1}$. 
Here we are interested in plasma wakefield excitations, which  produce a radiated field amplitude $E$ that scales as $E\simeq N_b\omega_{max}^2 \simeq N_b/\sigma_t^2$ , where $N_b$  is the number of particles in the beam.  
To maximize the wake amplitude, one utilizes, $\omega_{max}=\omega_p$, giving the wake optimization condition $\omega_p\sigma_t \simeq 1$. 
This scaling has been investigated experimentally \cite{hogan,blumenfeld}, theoretically and computationally \cite{rosen-prstab}, and its applicability even in the nonlinear ÒblowoutÓ regime \cite{rosen-pra-91} has been established. 
In fact, operation in the blowout regime is desired, as in this the plasma electrons are ejected from the beam channel, producing acceleration dependent only on longitudinal position $\xi=z-v_bt$  in the beam channel, where $v_b\simeq c$  is the beam  $z$-directed velocity. Further, the nominally uniform ion density in the electron-free region gives focusing linearly dependent on radial position  $r$ relative to the beam ($z$) axis. 
Thus the forces acting on the beam may produce high quality beams with the phase space density demanded by advanced applications in high-energy physics and light sources. 
The purpose of this Letter is to explore the physics implications of using this new class of  beam in the context of a plasma wakefield acceleration (PWFA) experiment, showing that it may yield fields of unprecedented amplitude and time scale, opening unique and compelling opportunities in plasma, accelerator, and atomic physics. 
We proceed by investigating, with the aid of simulations, the attributes of an optimized PWFA experiment using the LCLS 2~fs beam. We examine as well the attendant challenges in implementing such a scenario, in which the beam temporal scales and phase space qualities have as yet not been experimentally encountered.

In the case of the highly compressed LCLS beam, the optimization  $\omega_p\sigma_t\simeq 1$ yields a plasma electron density $n_0=\left[4\pi r_e (c\sigma_t)^2\right]^{-1}\simeq 7.8\times 10^{19}\mbox{cm}^{-3}$  corresponding to a gas density of several atm.  
Further, one may estimate whether this scenario may access the blowout regime, which requires that the beam density $n_b$  exceed $n_0$ , by evaluating the ratio of beam electrons to plasma electrons found within a volume of a cubic plasma skin-depth ($k_p^{-1}=c/\omega_p$), $\tilde{Q} \simeq N_b k_p^3/n_0=4\pi k_p r_e N_b$  \cite{rosen-prstab}. 
In our case we have $\tilde{Q} = 7.4$. 
Additionally, one must focus the beam to a rms size $\sigma_x$  smaller than a skin-depth, $k_p\sigma_x<1$, to enter the blowout regime ($n_b/n_0\gg$1). 
The equilibrium beam size in the plasma arising from ion focusing in the plasma-electron-free region is a function of $\epsilon_{n,x}$ and the equilibrium $\beta$-function, $\beta_{eq}=k_p^{-1}\sqrt{\gamma/2}$, as $\sigma_{x,eq}\simeq \sqrt{\beta_{eq}\epsilon_{n,x}/\gamma}$ . 
For the parameters above we have $\beta_{eq} = 140\mbox{$\mu$m}$ and, taking the maximum of the achieved $\epsilon_n$, $\sigma_{x,eq}=45\mbox{nm}$, and $k_p\sigma_{x,eq}=0.075$. 
The associated beam density is $n_b=6.5\times10^{21}\mbox{cm}^{-3}$, giving $n_b/n_0\simeq 84$, a highly nonlinear blowout scenario.  In such a case, one may expect the onset of effects such as ion motion. 

\begin{figure}[h]
\scalebox{.26}{\includegraphics[angle=-90]{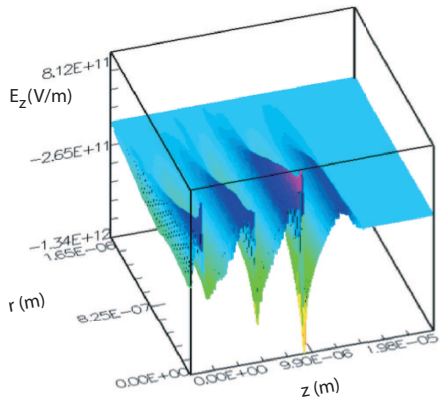}}
\caption{\label{fig:plasmawake}Longitudinal plasma wakefield for LCLS beam conditions described in text, $n_0=7.8\times10^{19}\mbox{cm}^{-3}$, and injected transverse beam size $\sigma_x=77$nm.}
\end{figure}

It will be seen that it is difficult and unnecessary to inject the beam into the plasma with $\beta=\beta_{eq}$; one must minimally create the conditions for plasma and wake formation, and then tolerate the beam size oscillations that result from having an initially mismatched $\beta$. 
Thus to gauge the strength of the plasma wakes in initial particle-in-cell (PIC) simulations we take the injected beam to have $\sigma_x\simeq 77 \mbox{nm} (n_b/n_0\simeq 28)$, corresponding to the average $\beta$ encountered in the plasma. These simulations were carried out using the 2D electro-magnetic PIC code OOPIC Pro \cite{oopic}, with the predictions for the longitudinal plasma wake $E_z(r,\xi)$ shown in Fig.~\ref{fig:plasmawake}. 
This scenario produces unprecedented large fields, in excess of 1.3~TV/m. 
Based on such accelerating gradients, one may envision a key component of a TeV-class collider existing on a table-top.  
Further, this field amplitude is remarkably large even by internal atomic physics standards. 
Indeed, the collective field of such beams can readily ionize matter, allowing both plasma formation, and a unique probe of atomic physics in an extreme field limit. 
The size of $E_z$ is not surprising, however, as the transverse vacuum $E$-field at beam edge for $\sigma_x=77$ nm can be estimated as $E_{r,max}\simeq eN_b/2(4\pi)^{1/2}\epsilon_0 \sigma_x \sigma_z = 1.5\mbox{TV/m}$, or approximately that of the induced longitudinal plasma wakefield field. 
With this scenario, we should restrict our interaction to $\sim$1mm, giving 
a maximum incurred energy loss of $<$1~GeV; in this case chromatic effects in transport do not cause beam loss in the experimentally sensitive region downstream of the LCLS undulator. 

We now examine the focusing needed to yield sufficiently small  $\sigma_{x(y)}$ that produces conditions for plasma creation and efficient wake formation. Further, to implement this experiment in the spatially-limited region  downstream of the LCLS undulator, we should make the focusing system compact (1.2~m in length) and easily implemented. 
As such, we have designed an ultra-high field gradient, permanent magnet quadrupole (PMQ) based focusing system similar to that developed for previous UCLA/LLNL experiments on inverse Compton scattering \cite{lim}. 
This short focal length system, which is based on quadrupole gradients of over 700~T/m, has the further advantage of evading performance restrictions due to chromatic aberrations. 
It satisfies space restrictions, and is simple and light-weight, giving the possibility of beamline insertion with minimum downtime. A collimating focusing system downstream of the spot waist, having reflection symmetry with the final focus system can be employed to capture and transport the beam exiting the plasma region. 
The final focus system design parameters are: beam energy 14.3 GeV, rms relative momentum spread $\sigma_{\delta p/p}=3\times10^{-5}$, initial $\beta$-function of 400~m, and $\epsilon_n=4\times10^{-7}\mbox{m-rad}$. 
With the PMQs arranged in a triplet configuration of 12.5, 25, and 25~cm magnet lengths, ELEGANT simulations indicate achievable beam sizes of $\sigma_x(y)=130\mbox{nm}$. 
This value is well suited for injection into the plasma, with a final $\beta=1.2\mbox{mm}$, a bit longer than the proposed plasma length.  

The amplitude of the beam's electric field at the tight focus is of high interest in its own right, as field levels that access the barrier suppression ionization (BSI) regime \cite{bauer} can be created in a wide variety of atomic species. 
This implies that the desired plasma density may be, as in previous SLAC FFTB experiments, created using the electron beam \cite{oconnell}.  
This has been the case in the FFTB scenarios, but with much longer beam time scales \cite{oconnell}, and with a factor of ~25 smaller fields, in which one attributes plasma formation to tunneling ionization.  There is uncertainty in the applicability of the theoretical models \cite{mevel} of field ionization, particularly in the case of ultra-short unipolar field pulses characteristic this type of charged particle beam, however. 
While we will employ such models \cite{oopic} to illustrate ultra-fast ionization for plasma creation below, this uncertainty argues forcefully for the performing of experiments.  

The self-consistent ionization of a high pressure gas by the beam fields must occur in 100's of as, to create the plasma quickly enough to give maximum amplitude plasma wakefields. 
Depending on the atomic species chosen, the fields at the front edge of the beam may yield ionization in the tunneling regime. Indeed, in OOPIC, a computational tool developed to describe ionization in the PWFA and related contexts, a tunneling model based on Ammosov-Delone-Krainov (ADK) theory \cite{adk} is employed, which is appropriate for fields in the region $>2\sigma_t$ ahead of beam center - the region of highest interest - in our case. 
In the current scenario, the much higher fields in the beam core provoke ionization, for most atomic species considered, in the BSI regime.  The boundary for the two regimes is delineated by a critical electric field value $E_{cr}$, which for hydrogen is $E_{cr}=439\mbox{~GV/m}$, while for Li, used in the previous generation of PWFA experiments, $E_{cr}=109\mbox{~GV/m}$. 

\begin{figure}[h]
\scalebox{.26}{\includegraphics[angle=-90]{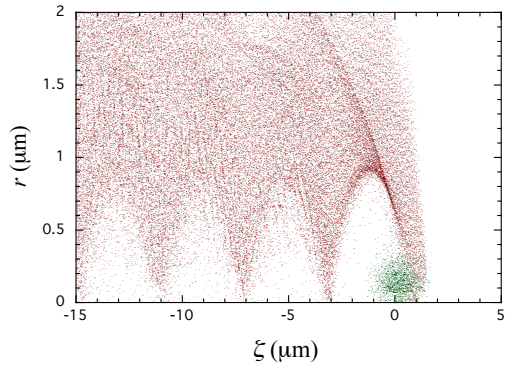}}
\caption{\label{fig:beamH2}Beam (green) and plasma (red) electrons in simulation of LCLS 20~pC beam initially focused to $\sigma_x = 130\mbox{~nm}$ passage through 3.15~atm H$_2$ gas.}
\end{figure}

Assuming the initial beam size  given by ELEGANT simulations, in the absence of plasma $E_{r,max}\simeq 800\mbox{~GV/m}$.  It is encouraging that, as seen in Figure~\ref{fig:beamH2}, full ionization based on the tunneling model in OOPIC is achieved using 3.15 atm H$_2$ gas (with a mm-width jet envisioned for use), and it occurs within the first several 100~as of the beam pulse.  This means that the plasma is created with appropriate $n_0$ well before the main portion of the beam passes, and longitudinal wake is nearly as in the pre-formed plasma case, $>$1.2~TV/m.   Further, even though the field amplitude indicates ionization in the BSI regime in the beam core, we need not be concerned with this region, and may ignore the lack of BSI in the simulations. 

To check the consistency of the OOPIC model of tunneling ionization with the transition to the BSI regime, we have used the formalism of Ref. \cite{bauer} to integrate the ionization rate equations including BSI for hydrogen, assuming a Gaussian-field impulse $\sigma_t=2\mbox{~fs}$ of 800~GV/m field amplitude.  
We find ionization occurring again well before ($\sim2\sigma_t$) pulse center. In this scenario, significant ionization rates in H due to BSI ($>\mbox{fs}^{-1}$) occur when the field exceeds $\sim$40~GV/m. In sum, we thus expect full tunneling ionization of H$_2$ gas in the rising edge of the beam, with possible assist from a BSI contribution to rates at higher fields. Of course, there is transverse field variation as well as an induced $E_z$ that should be included to fully understand the ionization behavior. This will be done in the simulation code in the future; the experiments then serve to benchmark theoretical and computational models in this newly accessible, extreme regime. 

In this regard, one sees that the LCLS 2~fs beam may be employed as an unprecedented method for probing the transition from tunneling ionization to BSI in a wide variety of atomic species, by scanning the value of $E_{r,max}$ through change in the focused beam size. 
This beam represents a tool that thus extends the use of half-cycle-type electromagnetic pulses from the study of ionization in Rydberg atoms \cite{jones} to those in the ground state.  
One may also use these beams in the vicinity of solid material to exploit their properties in exciting condensed matter, for investigating, e.g. magnetic switching \cite{tudosa}.

The generation of TV/m plasma wakefields means that one may measure beam energy changes with ease. 
Other experimental beam parameters present measurement challenges. 
With such short beams, diagnosing the pulse length requires an extension of experimental methods. The current fastest direct measurement of pulse lengths involves use of RF deflectors, but at present these systems lack the needed sub-fs resolution.  Optical-IR bunching in the FEL context has, on the other hand, been measured using coherent transition radiation (CTR) \cite{tremaine}. 
With CTR, the observed radiation spectrum is essentially that associated with the electron beamÕs temporal profile. 
One may also use non-destructive radiative methods such as coherent edge radiation (CER \cite{chubar}) to obtain a spectrum equivalent to that of CTR, as has been recently observed \cite{andonian}.  
In these cases, one may deduce bunch length and shape from single-shot measurement of the coherent radiation spectrum.

\begin{figure}[h]
\scalebox{.25}{\includegraphics[angle=-90]{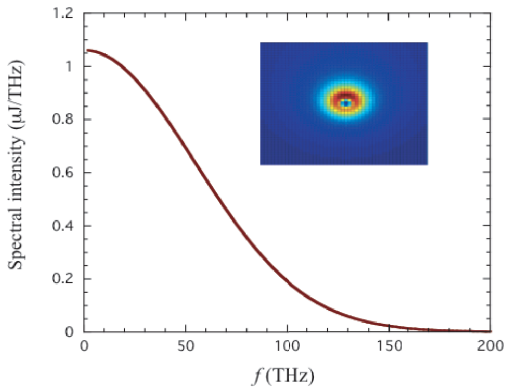}}
\caption{\label{fig:cer}Simulated coherent edge radiation spectrum from entrance into 1~T magnetic field, calculated by QUINDI for LCLS low charge beam parameters; (inset) far-field angular distribution.}
\end{figure}

In order to illustrate the expected signals in such a measurement, we have simulated, using a UCLA-developed Lenard-Wiechert field code termed QUINDI \cite{andonian}, the radiation spectra for CER emitted from an equivalent Gaussian-profile beam with LCLS parameters, arising during entrance into a 1 T dipole magnetic field. 
The result of this study shown in Figure~\ref{fig:cer}. 
The spectrum displayed is a Gaussian-like, with a cut-off frequency near  $f_c=(2\pi\sigma_t)^{-1}=83\mbox{~THz}$, as expected.  It reflects the beamÕs longitudinal Fourier spectrum because ER like TR is flat in its frequency response.  This similarity is also seen in the far-field angular distribution (inset in Figure~\ref{fig:cer}),  which displays a TR-like hollow pattern.  
We note that beyond its spectrum the emitted CER pulse is unique, as high power sub-cycle sources do not exist in this region of the IR.  
The pulse simulated in Fig.~\ref{fig:cer} has a peak power $>$200~MW, making it an ideal source for synchronized pumping of physical systems that can be probed by the coherent, fs X-ray pulses of the LCLS.

The measurement of few-100  $\sigma_x$ transverse sizes presents more formidable challenges.  One might consider optical transition radiation, OTR; we expect here also coherent OTR extending to near optical wavelengths, complicating the interpretation of such a measurement \cite{loos}.
An even larger difficulty is presented by the destructive nature of the focused beam to conducting surfaces used to create OTR.  On the other hand, one can exploit the collective beam fields to indirectly give the beam spot size, by measuring the ionization output of the beam passage through a gas. 
If several atomic species are used, the beam fields may be deduced, along with an estimation of concomitant $\sigma_x$. 
In the end, one does not need to focus more tightly than the spot that produces adequate ionization, $-$ sub-200 nm $-$ in the H case. 

\begin{figure}[h]
\scalebox{.26}{\includegraphics[angle=-90]{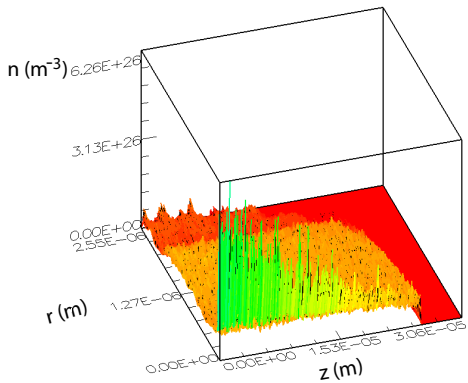}}
\caption{\label{fig:density}Ion density as a function of $r$ and $z$, with beam located as in same position in window as in Figure~\ref{fig:beamH2}, showing collapse of distribution (density spiking) behind the beam.}
\end{figure}

We note again that this injected beam, mismatched to the ion focusing strength, is initially focused in the plasma, reaching sizes $<\sigma_{x,eq}$, and then periodically oscillates with wavelength  $2\pi\beta_{eq}=0.88\mbox{~mm}$.
At the minimum $\sigma_x$, its density increases and the assumption of stationary ions can be violated. 
This effect causes the ions to collapse inside the beam, forming a region of nonuniform, higher ion density $n_i$. 
This produces in turn nonlinear fields that are potentially damaging to the extremely low
$\epsilon _n$ demanded by linear colliders \cite{rosen-prl-05}.  
This effect is not easily created in existing beams, due to the beam densities needed. 
In our case, we may enter into a regime of nontrivial ion motion in the current scenario, as illustrated by the OOPIC simulations in Fig.~\ref{fig:density}.  
The predicted peak $n_i$ is enhanced by a factor of 8 over nominal.  

However, as the ions do not move until after beam passage, they cannot in this case cause observable growth in $\epsilon _n$. 
This is desirable in terms of demonstrating the maintenance of beam quality in this extremely field scenario. 
On the other hand, the beam-derived forces produce ions having kinetic energy in the 1-100~kV range that may be collected to give a signature of ion motion.  

In conclusion, we have identified a compelling opportunity, based on the emergence of low charge, low emittance, few fs, high energy electron beams, to investigate an array of topics in high field physics. 
The primary motivation here is production of TV/m plasma wakefields, thus pushing the state-of-the-art by well over an order of magnitude, with dramatic implications on the size of future high-energy frontier accelerators. 
In analyzing the practical implementation of such an experiment at the LCLS, we find many aspects of the scenario that strongly add to its appeal. Most importantly, beams having peak space-charge fields on the order of a TV/m give an unprecedented tool for investigating the behavior of atomic systems under ultra-fast application of unipolar, very large amplitude electric fields.  
These fields also may allow exploration of potentially critical ion motion effects in a PWFA. 
Further, beams with these self-fields can yield prompt, coherent electromagnetic emission through a variety of mechanisms such as edge radiation,  thus creating a novel, high power source of half-cycle IR radiation not accessible by other means. 
This radiation thus serves as both a unique source and as a means of bunch length diagnosis. 
We have also discussed innovative approaches to diagnosis of $\sigma_x$, as well as beam transport and focusing and found that an experiment at the LCLS, having an appropriately small footprint in the post-undulator region, seems to be realizable. A collaboration based on the authorship of this Letter is now preparing a proposal to perform such an experiment.

\begin{acknowledgments}
\end{acknowledgments}

\bibliography{tevm2}

\end{document}